\documentclass[aps,prb,twocolumn]{revtex4-1}%
\usepackage{amsfonts}
\usepackage{amsmath}
\usepackage{amssymb}
\usepackage{graphicx}
\usepackage{txfonts}%
\providecommand{\U}[1]{\protect\rule{.1in}{.1in}}

\begin{document}
\title{Quantum measurement of hyperfine interaction in nitrogen-vacancy center}
\author{Kilhyun Bang}
\affiliation{Center for Advanced Nanoscience, Department of Physics, University of California San Diego, La Jolla, California 92093-0319, USA}
\author{Wen Yang}
\altaffiliation{Current address: Beijing Computational Science Research Center, Beijing 100084, China}
\affiliation{Center for Advanced Nanoscience, Department of Physics, University of California San Diego, La Jolla, California 92093-0319, USA}
\author{L. J. Sham}
\affiliation{Center for Advanced Nanoscience, Department of Physics, University of California San Diego, La Jolla, California 92093-0319, USA}

\pacs{PACS number}

\begin{abstract}
We propose an efficient quantum measurement protocol for the hyperfine
interaction between the electron spin and the $^{15}$N nuclear spin of a
diamond nitrogen-vacancy center. In this protocol, a sequence of quantum
operations of successively increasing duration is utilized to estimate the
hyperfine interaction with successively higher precision approaching the
quantum metrology limit.
This protocol does not need the preparation of the nuclear spin state. In the presence of realistic operation errors and electron spin decoherence, the overall precision of our protocol still surpasses the standard quantum limit.

\end{abstract}

\maketitle

\section{Introduction}

The negatively charged nitrogen-vacancy (NV) center in diamond is a promising
solid state system for quantum computation. The electron spin in the optical
ground state of the NV center exhibits exceptionally long coherence time
($>350$ $\mathrm{\mu}$s) at room temperature.\cite{Gaebel:2006fk} This
feature allows coherent manipulation and reliable readout of the state of the
electron spin and the neighboring nuclear spins
\cite{Neumann;Science;2010,Fuchs:2011fk} in the NV center, a key technique of
diamond-based quantum computation.
\cite{Childress:2006uq,Gaebel:2006fk,Neumann:2008fk,Fuchs:2011fk} In these
operations, the hyperfine interaction between the electron spin and the
neighboring nitrogen nuclear spin plays an important role. To minimize the
operation errors, an accurate estimate of the hyperfine interaction is desirable.

In addition to quantum computation, the NV center is also a candidate for the
application of quantum parameter estimation (also known as quantum metrology). Quantum metrology seeks quantum measurement protocols to
estimate physical parameters up to a given precision defined as $1/\Delta^{2}$ (with
$\Delta$ being the standard deviation) using the least amount $R$ of
resources, which include the number of measurements, the total duration of the
measurements, and the number of particles involved in the measurements. The
classical protocol utilizes the number $R$ of repeated measurements as a
resource and, according to the central limit theorem, gives the classical
limit (also known as standard quantum limit or SQL) $\Delta_{\mathrm{{SQL}}%
}=O(1/\sqrt{R})$. Quantum metrology aims to surpass the SQL and, more
ambitiously, reach the quantum metrology limit (QML) $\Delta_{\mathrm{{QML}}%
}=O(1/R)$, the upper precision bound $1/\Delta_{\mathrm{QML}}^{2}=O(R^{2})$
set by quantum mechanics. The most popular quantum measurement technique is
interferometry, in which the parameter to be measured is recorded as a phase
in the coherence of the system.\cite{Giovannetti19112004,Giovannetti;PRL;2006,
Dunningham;ContemPhys;2006,Dowling} The exceptionally long coherence time of
the NV center electron spin diminishes the detrimental effect of decoherence
on such measurements and makes the NV center an ideal system for quantum
metrology.\cite{Goldstein;PRL;2011} Up to date, most of the measurement
protocols utilize pure quantum states and surpass the SQL by creating quantum
entanglement in the system. However, the thermal equilibrium state of the
nuclear spins is highly mixed at room temperature. To estimate \textit{reliably }the hyperfine interaction in the NV
center by a pure-state protocol, the nuclear spins must be prepared repeatedly
into a given pure state. Further, the number of spins as the resources of
entanglement in a single NV center is finite,\cite{Neumann:2008fk} so the
advantage of quantum entanglement to parameter estimation is also limited.

Recently, Boixo and Somma\cite{Boixo;PRA;2008} proposed a model of mixed-state
quantum metrology by combining the mixed-state quantum computation (also known
as deterministic quantum computation with one quantum bit\cite{Knill;PRL;1998}
or DQC1) with the adaptive Bayesian inference. This DQC1 model utilizes the
total duration $T$ (instead of large-scale
entanglement\cite{Goldstein;PRL;2011}) of the estimation process as a resource
to approach the QML $\Delta_{\mathrm{{QML}}}=O(1/T)$ without creating any
entanglement.\cite{Datta;PRL;2008,Lanyon;PRL;2008} However, its application
to estimate the hyperfine interaction in the NV center requires including the effects of noise and unintended dynamics.

In this paper, we construct an efficient quantum measurement protocol to
estimate the hyperfine interaction between the electron spin and the $^{15}$N
nuclear spin in the NV center. This protocol is essentially a combination of
the DQC1 model\cite{Boixo;PRA;2008} and the spin-echo technique,\cite{Yang;FP;2011} which
decouples the dynamics driven by the hyperfine interaction from the noise and
unintended dynamics. It does not need the preparation of the nuclear spin
state and approaches the QML $\Delta_{\mathrm{{QML}}}=O(1/T)$ in the ideal
case. By including realistic errors (such as the nuclear spin rotation error
and the electron spin decoherence) in our analysis, we show that our
protocol still surpasses the SQL under typical experimental conditions.

The rest of this paper is organized as follows. In Sec.~\ref{SEC_DQC1}, we
review the DQC1 model for parameter estimation and identify the problems in
applying this model to estimate the hyperfine interaction in the NV
center. In Sec.~\ref{SEC_SOLUTION}, we give a solution to these problems by
combining the DQC1 model with the spin-echo technique.
In Sec.~\ref{SEC_OURPROTOCOL}, we introduce our quantum measurement
protocol. Sec.~\ref{SEC_CONCLUSION} gives the conclusion.

\section{DQC1 parameter estimation in NV center}

\label{SEC_DQC1}

We first review the two-qubit version of the DQC1 parameter
estimation model proposed by Boixo and Somma\cite{Boixo;PRA;2008} (Sec.~\ref{SEC_DQC1_1}) and then identify the robustness problems arising from
applying this model to estimate the hyperfine interaction in the NV center (Sec.~\ref{SEC_DQC1_2}).

\subsection{Two-qubit DQC1 parameter estimation}

\label{SEC_DQC1_1}

\begin{figure}[ptb]
\includegraphics[width=0.9\columnwidth,clip]{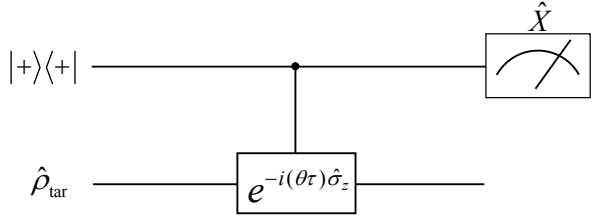}\caption{DQC1
parameter estimation with one control qubit in the pure state $\left\vert
+\right\rangle \equiv(\left\vert 0\right\rangle +\left\vert 1\right\rangle
)/\sqrt{2}$ and one target qubit in the state $\hat{\rho}_{\mathrm{tar}}$.}%
\label{DQC1}%
\end{figure}

The two-qubit DQC1 model consists of a control qubit (with states
$\{\left\vert 0\right\rangle ,\left\vert 1\right\rangle \}$) and a target
qubit (with states $\{\left\vert \uparrow\right\rangle ,\left\vert
\downarrow\right\rangle \}$). The initial state $\hat{\rho}_{\mathrm{DQC1}%
}=\left\vert +\right\rangle \left\langle +\right\vert \otimes\hat{\rho
}_{\mathrm{tar}}$ is the direct product of the pure state $\left\vert
+\right\rangle \equiv(\left\vert 0\right\rangle +\left\vert 1\right\rangle
)/\sqrt{2}$ of the control qubit and the unpolarized state $\hat{\rho
}_{\mathrm{tar}}=(\left\vert \uparrow\right\rangle \left\langle \uparrow
\right\vert +\left\vert \downarrow\right\rangle \left\langle \downarrow
\right\vert )/2$ of the target qubit, as shown in Fig.~\ref{DQC1}.
The three Pauli operators of the control qubit and of the target qubit are denoted by
\begin{align*}
\hat{X}  &  \equiv\left\vert 1\right\rangle \left\langle 0\right\vert
+\left\vert 0\right\rangle \left\langle 1\right\vert ,\\
\hat{Y}  &  \equiv i(\left\vert 1\right\rangle \left\langle 0\right\vert
-\left\vert 0\right\rangle \left\langle 1\right\vert ),\\
\hat{Z}  &  \equiv\left\vert 0\right\rangle \left\langle 0\right\vert
-\left\vert 1\right\rangle \left\langle 1\right\vert ,
\end{align*}
and $\{\hat{\sigma}_{x},\hat{\sigma}_{y},\hat{\sigma}_{z}\}$, respectively.
The two qubits are coupled by the interaction
\begin{equation}
\hat{H}_{\mathrm{DQC1}}=\left\vert 1\right\rangle \left\langle 1\right\vert
\otimes\theta\hat{\sigma}_{z}. \label{HI_DQC1}%
\end{equation}
This interaction makes the splitting energy $\omega_{c}$ of the control qubit
dependent on the state of the target qubit: $\omega_{c,\uparrow}=\theta$ for
the target qubit in the spin-up state $\left\vert \uparrow\right\rangle $ and
$\omega_{c,\downarrow}=-\theta$ for the target qubit in the spin-down state
$\left\vert \downarrow\right\rangle $. The DQC1 parameter
estimation\cite{Boixo;PRA;2008} aims to estimate the interaction strength
$\theta$ with the standard deviation $\Delta_{\theta}=O(1/T)$ approaching the
QML, where $T$ is the total duration of the estimation process. The procedures
are simple: the application of the two-qubit interaction $\hat
{H}_{\mathrm{DQC1}}$ for a duration $\tau$, followed by a measurement of
$\hat{X}$:

\begin{itemize}
\item If the target qubit is in the spin-up state $\left\vert \uparrow
\right\rangle $, then $\hat{H}_{\mathrm{DQC1}}$ drives the precession of the
control qubit with angular frequency $\omega_{c,\uparrow}$,
\[
\frac{\left\vert 0\right\rangle +\left\vert 1\right\rangle }{\sqrt{2}}%
\otimes\left\vert \uparrow\right\rangle \rightarrow\frac{\left\vert
0\right\rangle +e^{-i\omega_{c,\uparrow}\tau}\left\vert 1\right\rangle }%
{\sqrt{2}}\otimes\left\vert \uparrow\right\rangle .
\]
Before the measurement, the interaction strength $\theta$ is encoded as a
phase $e^{-i\omega_{c,\uparrow}\tau}$ of the control qubit. The repeated
measurements of $\hat{X}$ estimate the average value $\langle\hat{X}%
\rangle_{\uparrow}=\cos(\omega_{c,\uparrow}\tau)=\cos(\theta\tau)$, which
yields the phase.

\item If the target qubit is in the spin-down state $\left\vert \downarrow
\right\rangle $, then $\hat{H}_{\mathrm{DQC1}}$ drives the precession of the
control qubit with angular frequency $\omega_{c,\downarrow}$,
\[
\frac{\left\vert 0\right\rangle +\left\vert 1\right\rangle }{\sqrt{2}}%
\otimes\left\vert \downarrow\right\rangle \rightarrow\frac{\left\vert
0\right\rangle +e^{-i\omega_{c,\downarrow}\tau}\left\vert 1\right\rangle
}{\sqrt{2}}\otimes\left\vert \downarrow\right\rangle .
\]
Before the measurement, the interaction strength $\theta$ is encoded as a
phase $e^{-i\omega_{c,\downarrow}\tau}$ of the control qubit. The repeated
measurements of $\hat{X}$ estimate the average value $\langle\hat{X}%
\rangle_{\downarrow}=\cos(\omega_{c,\downarrow}\tau)=\cos(\theta\tau)$, which
extracts the phase.

\item Now the target qubit is in the unpolarized state, i.e., an equal,
incoherent mixture of $\left\vert \uparrow\right\rangle $ and $\left\vert
\downarrow\right\rangle $. Then the repeated measurements of $\hat{X}$
estimates the equally weighted average of $\langle\hat{X}\rangle_{\uparrow}$
and $\langle\hat{X}\rangle_{\downarrow}$:
\[
\langle\hat{X}\rangle=\frac{1}{2}(\langle\hat{X}\rangle_{\uparrow}+\langle
\hat{X}\rangle_{\downarrow})=\cos(\theta\tau).
\]
A distinctive feature of the above parameter estimation process is the absence
of any two-qubit entanglement.\cite{Lanyon;PRL;2008}
\end{itemize}

For a given standard deviation $\Delta_{X}$ ($\ll1$ under typical situations)
in estimating $\langle\hat{X}\rangle$, the DQC1 model gives an estimate to the
interaction strength $\theta$ with a standard deviation%
\begin{equation}
\Delta_{\theta}=\frac{\Delta_{X}}{|\partial\langle\hat{X}\rangle
/\partial\theta|}=\frac{\Delta_{X}}{\tau|\sin(\theta\tau)|}\geq\frac
{\Delta_{X}}{\tau}. \label{DELTA}%
\end{equation}
By regarding the duration $\tau$ of the estimation as a resource, the QML scaling
$\Delta_{\theta}=O(1/\tau)$ is achieved if $\tau$ could be chosen such that
$|\sin(\theta\tau)|\approx1$. However, due to the limited prior knowledge
about $\theta$ (the parameter to be estimated), we cannot always ensure
$|\sin(\theta\tau)|\approx1$, especially when a small standard deviation
$\Delta_{\theta}\rightarrow0$ (corresponding to large $\tau\rightarrow\infty$)
is required.

To address this issue, Boixo and Somma\cite{Boixo;PRA;2008} quantified the prior
knowledge about $\theta$ by a standard deviation $\Delta_{0}$ and utilized the
adaptive Bayesian inference to reduce the standard deviation successively. The
essential idea of this approach can be understood \textit{qualitatively }as
follows. In order to ensure $|\sin(\theta\tau)|\approx1$ and hence the QML,
the largest $\tau$ is roughly $1/\Delta_{0}$. Under this restriction, the
minimal standard deviation for the estimation of $\theta$ is given by
Eq.~(\ref{DELTA}) as $\sim\Delta_{X}\Delta_{0}\ll\Delta_{0}$. Therefore, the
DQC1 measurements with standard deviation $\Delta_{X}$ refines our knowledge
about the interaction strength $\theta$ from a large standard deviation
$\Delta_{0}$ to a much smaller one $\sim\Delta_{X}\Delta_{0}$. By iterating
this procedure, the standard deviation $\Delta_{\theta}$ would decrease
successively as $\Delta_{0}\rightarrow\Delta_{X}\Delta_{0}\rightarrow
\Delta_{X}^{2}\Delta_{0}\rightarrow\cdots$. With the aid of the adaptive
Bayesian inference, Boixo and Somma\cite{Boixo;PRA;2008} performed a
quantitative analysis about this iteration and concluded that the QML
$\Delta_{\theta}=O(1/T)$ could be achieved for an arbitrary desired standard
deviation, where $T=\sum\tau$ is the total duration of the estimation process.

In the next subsection, we discuss the problems of DQC1 model when it is
directly applied to estimate the hyperfine interaction in the NV center.
Before that, we mention a useful extension (which can be readily
verified) of this model: the analytical expressions for the quantity estimated
by the measurement [e.g., $\langle\hat{X}\rangle=\cos(\theta\tau)$ for the
DQC1 model and $\langle\hat{Z}\rangle=\cos(A\tau)$ for our protocol, see Eq.~(\ref{Zop})] remains valid for a more general initial state $\rho
_{\mathrm{tar}}=1/2+q_{z}\sigma_{z}/2$ of the target qubit with an arbitrary
polarization $q_{z}$. This fact is especially important for estimating the
hyperfine interaction in the NV center since in this case, initializing the
control qubit (the electron spin in the NV center) will partially polarize the
target qubit (the $^{15}$N nuclear spin in the NV
center).\cite{Jacques;PRL;2009}

\subsection{Direct application of DQC1 parameter estimation to NV center}

\label{SEC_DQC1_2}

\begin{figure}[ptb]
\includegraphics[width=\columnwidth,clip]{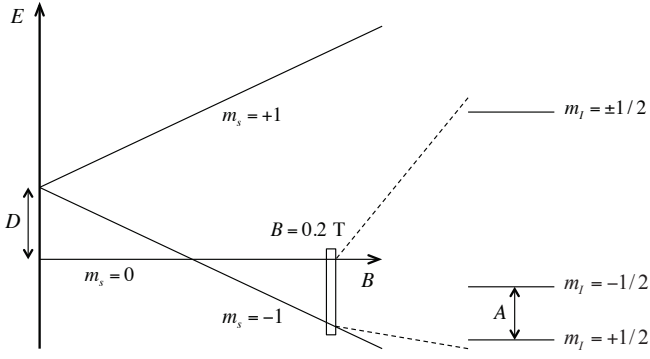}\caption{Energy
level diagram of the ground state of an NV center in diamond. The hyperfine
energy splitting at $B=0.2\text{ T}$ is sketched within the $\left\vert
m_{s}=0\right\rangle $ and $\left\vert m_{s}=-1\right\rangle $ manifold.
$D=2.87\text{ GHz}$ is the zero field splitting of the electron spin, and $A$
is the longitudinal hyperfine interaction to be estimated. The nuclear Zeeman
splitting is omitted in the diagram.}%
\label{NVenergy}%
\end{figure}

We consider a negatively charged NV center in diamond consisting of a
substitutional $^{15}\text{N}$ atom and a neighboring carbon vacancy. Its
electronic ground state is a two-electron spin triplet described by a spin-1
operator $\mathbf{\hat{S}}$, with a zero-field splitting $D\approx2.87$
$\mathrm{GHz}$ (described by the term $D\hat{S}_{z}^{2}$) between the
$\left\vert m_{s}=0\right\rangle $ state and the $\left\vert m_{s}%
=\pm1\right\rangle $ states. Under an external magnetic field $B$ along the
N-V axis (defined as the $z$ direction), the Zeeman term $g_{e}\mu_{B}B\hat
{S}_{z}$ with $g_{e}=2.0023$ shifts the state $\left\vert m_{s}%
=+1\right\rangle $ away from the other two states under a moderate
magnetic field $B\sim0.2$ T (see Fig.~\ref{NVenergy}). Thus we identify
$\left\vert 0\right\rangle \equiv\left\vert m_{s}=0\right\rangle $ and
$\left\vert 1\right\rangle \equiv\left\vert m_{s}=-1\right\rangle $ as the two
states of the control qubit of the DQC1 model and use $\hat{X},\hat{Y},\hat
{Z}$ as the three Pauli matrices for this qubit. The electron spin
$\mathbf{\hat{S}}$ is coupled to the neighboring $^{15}$N nuclear spin-1/2
$\mathbf{\hat{I}}$ (with the two-fold degeneracy lifted by the Zeeman term
$g_{N}\mu_{N}B\hat{I}_{z}$, where $g_{N}=-0.5664$\cite{Felton;PRB;2009})
through the hyperfine interaction $A\hat{S}_{z}\hat{I}_{z}+(A_{\perp}%
/2)(\hat{S}_{+}\hat{I}_{-}+\hat{S}_{-}\hat{I}_{+})$, where $A\approx3.03\text{
MHz}$ and $A_{\perp}\approx3.65\text{ MHz}$.\cite{Felton;PRB;2009} We regard
this nuclear spin-1/2 as the mixed-state target qubit of the DQC1 model and
use $\hat{\sigma}_{x},\hat{\sigma}_{y},\hat{\sigma}_{z}$ as the three Pauli
matrices $2\hat{I}_{x},2\hat{I}_{y},2\hat{I}_{z}$ for this qubit. The diagonal
part $A\hat{S}_{z}\hat{I}_{z}$ of the hyperfine interaction makes the nuclear
(electron) spin splitting energy dependent on the state of the electron (the
nucleus). Thus $A\hat{S}_{z}\hat{I}_{z}$ plays the central role in coherent
control and readout of the electron and nuclear spin states. The hyperfine
interaction strength $A$ is the parameter to be estimated.

In the two-qubit subspace, the Hamiltonian $\hat{H}=\hat{H}_{0}+\hat
{H}_{\mathrm{mix}}$ consists of the diagonal part
\[
\hat{H}_{0}=\frac{1}{2}g_{N}\mu_{N}B\hat{\sigma}_{z}+\left\vert 1\right\rangle
\left\langle 1\right\vert \otimes(D^{\prime}-\frac{1}{2}A\hat{\sigma}_{z})
\]
and the off-diagonal part
\[
\hat{H}_{\mathrm{mix}}=(A_{\perp}/\sqrt{2})(\left\vert 0,\downarrow
\right\rangle \left\langle 1,\uparrow\right\vert +\left\vert 1,\uparrow
\right\rangle \left\langle 0,\downarrow\right\vert ).
\]
The diagonal part $\hat{H}_{0}$ accounts for the free nuclear spin precession
with angular frequency $g_{N}\mu_{N}B$, the free electron spin precession with
angular frequency $D^{\prime}\equiv D-g_{e}\mu_{B}B$, and the projection
$\left\vert 1\right\rangle \left\langle 1\right\vert \otimes(-A\hat{\sigma
}_{z}/2)$ of the diagonal hyperfine interaction $A\hat{S}_{z}\hat{I}_{z}$ in
the two-qubit subspace. The off-diagonal part $\hat{H}_{\mathrm{mix}}$ is the
projection of the off-diagonal hyperfine interaction $(A_{\perp}/2)(\hat
{S}_{+}\hat{I}_{-}+\hat{S}_{-}\hat{I}_{+})$ in the two-qubit subspace. The
diagonal hyperfine interaction term $\left\vert 1\right\rangle \left\langle
1\right\vert \otimes(-A\hat{\sigma}_{z}/2)$ in $\hat{H}_{0}$ corresponds to
$\hat{H}_{\mathrm{DQC1}}$ in Eq.~(\ref{HI_DQC1}) with $\theta\leftrightarrow
(-A/2)$. It makes the precession frequency $\omega_{e}$ of the electron spin
dependent on the hyperfine interaction strength $A$ and the nuclear spin
state: $\omega_{e,\uparrow}=D^{\prime}-A/2$ for the nuclear spin state being
$\left\vert \uparrow\right\rangle $ and $\omega_{e,\downarrow}=D^{\prime}+A/2$
for the nuclear spin state being $\left\vert \downarrow\right\rangle $.
Therefore, following the procedure in Fig.~\ref{DQC1}, the interaction
strength $A$ is encoded as a phase of the electron spin and subsequently
extracted by estimating $\langle\hat{X}\rangle$.

\begin{figure}[ptb]
\includegraphics[width=\columnwidth,clip]{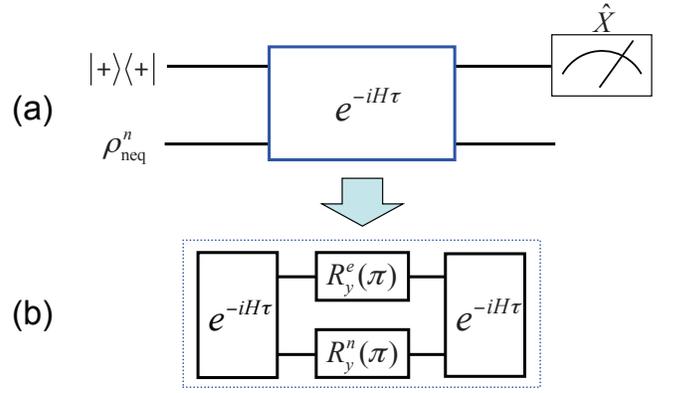}\caption{(a) Direct
application of the DQC1 model to estimate the hyperfine interaction strength
$A$ in NV center. (b) Combination of spin echo and the DQC1 model.}%
\label{G_DECOUPLING}%
\end{figure}

As schematically shown in Fig.~\ref{G_DECOUPLING}(a), the electron spin needs
to be prepared in the superposition $(\left\vert 0\right\rangle +\left\vert
1\right\rangle )/\sqrt{2}$. This can be achieved by optical pumping
\cite{Harrison:2004uq} followed by a coherent rotation. However, this
preparation process inevitably influences the nuclear spin and changes its
state from the unpolarized thermal equilibrium state $\hat{\rho}_{\mathrm{eq}%
}^{n}=\hat{I}/2$ to a state $\hat{\rho}_{\mathrm{neq}}^{n}=\hat{I}/2+q_{z}\hat{\sigma}%
_{z}/2$ with a finite polarization $q_{z}=\operatorname*{Tr}[\hat{\rho
}_{\mathrm{neq}}^{n}\hat{\sigma}_{z}]$.\cite{Jacques;PRL;2009} Then the two
qubits evolve under the Hamiltonian $\hat{H}$ for a duration $\tau$, followed
by a measurement of $\langle\hat{X}\rangle$. Below we calculate $\langle
\hat{X}\rangle$ without $\hat{H}_{\mathrm{mix}}$ and then taking it
into account by perturbation theory.

Without $\hat{H}_{\mathrm{mix}}$, the two qubits are driven by $\hat
{H}_{0}$, which has four eigenstates $\left\vert 0,\uparrow\right\rangle
,\left\vert 0,\downarrow\right\rangle ,\left\vert 1,\uparrow\right\rangle
,\left\vert 1,\downarrow\right\rangle $. The physics is similar to the DQC1
model described in the previous subsection:

\begin{itemize}
\item If the nuclear spin is in the spin-up state $\left\vert \uparrow
\right\rangle $, then $\hat{H}_{0}$ drives the precession of the electron spin
qubit with angular frequency $\omega_{e,\uparrow}$ and the repeated
measurements of $\hat{X}$ estimate $\langle\hat{X}\rangle_{\uparrow}%
=\cos(\omega_{e,\uparrow}\tau)$.

\item If the nuclear spin is in the spin-down state $\left\vert \downarrow
\right\rangle $, then $\hat{H}_{0}$ drives the precession of the electron spin
qubit with angular frequency $\omega_{e,\downarrow}$ and the repeated
measurements of $\hat{X}$ estimate $\langle\hat{X}\rangle_{\downarrow}%
=\cos(\omega_{e,\downarrow}\tau)$.

\item Now the nuclear spin is in an incoherent mixture of $\left\vert
\uparrow\right\rangle $ [with weight $(1+q_{z})/2$] and $\left\vert
\downarrow\right\rangle $ [with weight $(1-q_{z})/2$]. Then the repeated
measurements of $\hat{X}$ estimate the weighted average of $\langle\hat
{X}\rangle_{\uparrow}$ and $\langle\hat{X}\rangle_{\downarrow}$:%
\begin{equation}
\langle\hat{X}\rangle=\frac{1+q_{z}}{2}\langle\hat{X}\rangle_{\uparrow}%
+\frac{1-q_{z}}{2}\langle\hat{X}\rangle_{\downarrow}. \label{AVEX_AP0}%
\end{equation}

\end{itemize}

Then we consider the complications caused by the off-diagonal part $\hat{H}_{\mathrm{mix}}$. To reduce
its detrimental effect on the parameter estimation, we consider a suitable
magnetic field strength (e.g., $B=0.2$ T, as indicated in Fig. \ref{NVenergy}
and used in our estimation, see Sec. \ref{SEC_IDEAL}) so that $|D^{\prime}%
|\gg|A_{\perp}|$. In this case, we can use perturbation theory to treat
$\hat{H}_{\mathrm{mix}}$, which modifies the eigenstates and eigenenergies of
the two-qubit Hamiltonian $\hat{H}=\hat{H}_{0}+\hat{H}_{\mathrm{mix}}$:

\begin{enumerate}
\item $\hat{H}_{\mathrm{mix}}$ changes the eigenstates of $\hat{H}$ from
$[\left\vert 0,\uparrow\right\rangle $, $\left\vert 0,\downarrow\right\rangle
$, $\left\vert 1,\uparrow\right\rangle $, $\left\vert 1,\downarrow
\right\rangle ]$ to $[\left\vert 0,\uparrow\right\rangle $, $\widetilde
{\left\vert 0,\downarrow\right\rangle }$, $\widetilde{\left\vert
1,\uparrow\right\rangle }$, $\left\vert 1,\downarrow\right\rangle ]$, where%
\begin{align*}
\widetilde{\left\vert 0,\downarrow\right\rangle }  &  =[1-O(\eta
^{2})]\left\vert 0,\downarrow\right\rangle +O(\eta)\left\vert 1,\uparrow
\right\rangle ,\\
\widetilde{\left\vert 1,\uparrow\right\rangle }  &  =[1-O(\eta^{2})]\left\vert
1,\uparrow\right\rangle +O(\eta)\left\vert 0,\downarrow\right\rangle ,
\end{align*}
and $\eta\equiv A_{\perp}/(D^{\prime}+g_{N}\mu_{N}B-A/2)\sim10^{-3}$ for
$B=0.2$ T. In other words, $\hat{H}_{\mathrm{mix}}$ introduces new $O(\eta)$
components into the eigenstates. It can be readily verified that this changes
$\langle\hat{X}\rangle$ by $O(\eta^{2})$.

\item $\hat{H}_{\mathrm{mix}}$ changes the eigenenergy of $\left\vert
0,\downarrow\right\rangle $ ($\left\vert 1,\uparrow\right\rangle $) by a small
amount $-\delta$ ($+\delta$), where $\delta=\eta A_{\perp}/2+O(\eta
^{2}A_{\perp})$. This in turn changes the precession frequencies of the
electron spin from $\omega_{e,\mu}$ to $\tilde{\omega}_{e,\mu}=\omega_{e,\mu
}+\delta$ ($\mu=\uparrow,\downarrow$). Therefore, the average value
$\langle\hat{X}\rangle$ is obtained from Eq.~(\ref{AVEX_AP0}) by renormalizing
$\omega_{e,\mu}$ with $\tilde{\omega}_{e,\mu}$ $(\mu=\uparrow,\downarrow)$.
\end{enumerate}

Collecting both corrections discussed above, we obtain%
\begin{align}
\langle\hat{X}\rangle &  =\cos[(D^{\prime}+\delta)\tau]\cos(\frac{A}{2}%
\tau)\label{X}\\
&  +q_{z}\sin[(D^{\prime}+\delta)\tau]\sin(\frac{A}{2}\tau)+O(\eta
^{2}).\nonumber
\end{align}
It contains not only $A$ but also undesired parameters such as $D^{\prime}$
(free electron spin precession frequency), $\delta$ (energy shift by $\hat
{H}_{\mathrm{mix}}$), and $q_{z}$ (partial nuclear spin polarization). For an
accurate estimation of $A$, it is desirable to eliminate these undesired
parameters from $\langle\hat{X}\rangle$ by modifying the DQC1 protocol.

\section{Eliminating undesired parameters by spin echo}

\label{SEC_SOLUTION}

To remove the dependence on the undesired parameters in $\langle\hat{X}%
\rangle$, we combine the DQC1 model with the spin-echo technique by replacing
the free evolution $e^{-i\hat{H}\tau}$ with the composite evolution [see
Fig.~\ref{G_DECOUPLING}(b)]
\[
\hat{U}_{\mathrm{com}}=e^{-i\hat{H}\tau}\hat{R}_{y}^{n}(\pi)\hat{R}_{y}^{e}(\pi
)e^{-i\hat{H}\tau}=\hat{R}_{y}^{e}(\pi)\hat{R}_{y}^{n}(\pi)\left(  e^{-i(\hat{\sigma}_{y}\hat{Y}\hat{H}\hat{Y}\hat{\sigma}_{y})\tau}e^{-i\hat{H}\tau}\right)  ,
\]
which consists of an electron spin $\pi$ rotation $\hat{R}_{y}^{e}(\pi)=e^{-i\pi
\hat{Y}/2}=-i\hat{Y}$ and a nuclear spin $\pi$ rotation $\hat{R}_{y}^{n}%
(\pi)=e^{-i\pi\hat{\sigma}_{y}/2}=-i\hat{\sigma}_{y}$ sandwiched by the free
evolution $e^{-i\hat{H}\tau}$. This composite evolution contains a spin echo
(the part inside the parenthesis) for the electron and the nucleus, which
eliminates the free precession of the electron spin and the nuclear spin. To
analyze $\hat{U}_{\mathrm{com}}$ in more detail, we first ignore the
off-diagonal part $\hat{H}_{\mathrm{mix}}$ and then take it into account by
perturbation theory.

Without $\hat{H}_{\mathrm{mix}}$, the Hamiltonian $\hat{H}^{\prime}%
\equiv \hat{\sigma}_{y}\hat{Y}\hat{H}\hat{Y}\hat{\sigma}_{y}$ commutes with $\hat{H}$. Thus $\hat
{U}_{\mathrm{com}}$ reduces to
\[
\hat{U}_{\mathrm{com}}^{(0)}=\hat{R}_{y}^{e}(\pi)\hat{R}_{y}^{n}(\pi)e^{-i(\hat{H}%
^{\prime}+\hat{H})\tau}=\hat{R}_{y}^{e}(\pi)\hat{R}_{y}^{n}(\pi)e^{-iA\tau\hat{\sigma}_{z}/2}e^{-i\hat{H}_{\mathrm{echo}}\tau},
\]
where $\hat{H}_{\mathrm{echo}}=\left\vert 1\right\rangle \left\langle
1\right\vert \otimes(-A\hat{\sigma}_{z})$ corresponds to $\hat{H}%
_{\mathrm{DQC1}}$ in Eq.~(\ref{HI_DQC1}) with $\theta\leftrightarrow-A$. The
operation $\hat{R}_{y}^{n}(\pi)e^{-iA\tau\hat{\sigma}_{z}/2}$ on the nuclear spin
alone can be dropped since it does not influence our measurement on the
electron spin. Therefore, the composite evolution becomes $\hat{U}%
_{\mathrm{com}}^{(0)}=\hat{R}_{y}^{e}(\pi)e^{-i\hat{H}_{\mathrm{echo}}\tau}$, in
which all the undesired parameters have been eliminated.

In the presence of $\hat{H}_{\mathrm{mix}}$, $\hat{H}^{\prime}$ consists of the diagonal part
\[
\hat{H}_{0}^{\prime}=-\frac{1}{2}g_{N}\mu_{N}B\hat{\sigma}_{z}+\frac{1}%
{2}A\hat{\sigma}_{z}-\left\vert 1\right\rangle \left\langle 1\right\vert
\otimes(D^{\prime}+\frac{1}{2}A\hat{\sigma}_{z})
\]
and the off-diagonal part $\hat{H}_{\mathrm{mix}}$. Similar to the two-step
analysis leading to Eq.~(\ref{X}), $\hat{H}_{\mathrm{mix}}$
modifies the eigenstates and eigenenergies of $\hat{H}=\hat{H}_{0}+\hat
{H}_{\mathrm{mix}}$ and $\hat{H}^{\prime}=\hat{H}_{0}^{\prime}+\hat
{H}_{\mathrm{mix}}$:

\begin{enumerate}
\item $\hat{H}_{\mathrm{mix}}$ introduces new $O(\eta)$ components into the
eigenstates of $\hat{H}$ and $\hat{H}^{\prime}$. This changes $\langle\hat{X}\rangle$
by $O(\eta^{2})$.

\item For the Hamiltonian $\hat{H}$, the presence of $\hat{H}_{\mathrm{mix}}$
changes the eigenenergy of $\left\vert 0,\downarrow\right\rangle $
($\left\vert 1,\uparrow\right\rangle $) by $-\delta$ $(+\delta$). For the
Hamiltonian $\hat{H}^{\prime}$, the presence of $\hat{H}_{\mathrm{mix}}$
changes the eigenenergy of $\left\vert 0,\downarrow\right\rangle $
($\left\vert 1,\uparrow\right\rangle $) by $+\delta$ ($-\delta)$. In other words,
the opposite energy shifts for $\hat{H}$ and $\hat{H}^{\prime}$ induced by
$\hat{H}_{\mathrm{mix}}$ cancel each other in the evolution
$\hat{U}_{\mathrm{com}}$.
\end{enumerate}

For $\langle\hat{X}\rangle$, the composite evolution including both corrections
discussed above is equivalent to%
\[
\hat{U}_{\mathrm{com}}=\hat{R}_{y}^{e}(\pi)e^{-i\hat{H}_{\mathrm{echo}}\tau}%
+O(\eta^{2}),
\]
i.e., the spin echo eliminates all the named undesired parameters and the effective
evolution $\hat{U}_{\mathrm{com}}$ for the NV center recovers the DQC1
evolution $e^{-i\hat{H}_{\mathrm{DQC1}}\tau}$ up to a trivial electron spin
$\pi$ rotation $\hat{R}_{y}^{e}(\pi)$.

\section{Quantum measurement protocol of hyperfine interaction}

\label{SEC_OURPROTOCOL}

In this section, first we give the quantum circuit for a
single estimation of the hyperfine interaction strength $A$ in the NV center.
Second, we describe in detail the procedure of the entire estimation
protocol:\ the successive adaptation of the quantum circuit for dramatically
reduced standard deviation by combining our prior knowledge with the outcomes
of the previous measurements through adaptive Bayesian inference. Third, we
demonstrate that this protocol approaches the QML $\Delta_{\mathrm{{QML}}%
}=O(1/T)$ for the ideal case. Finally, we include the essential errors (the
nuclear spin rotation error and the electron spin decoherence) and show that
our protocol still exceeds the SQL.

\subsection{Quantum estimation circuit}

\label{SEC_CIRCUIT}

\begin{figure}[ptb]
\includegraphics[width=\columnwidth,clip]{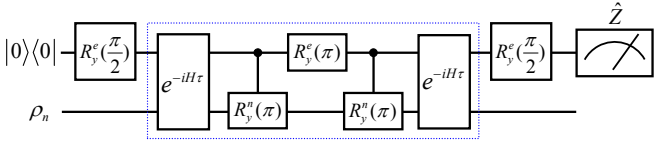}\caption{Quantum
circuit for a single estimation of the hyperfine interaction strength $A$ in
the NV center.}%
\label{G_PROTOCOL}%
\end{figure}

Fig.~\ref{G_PROTOCOL} gives the sequence of quantum operations for a single
estimation of the hyperfine interaction strength $A$ in the NV center:

\begin{enumerate}
\item The electron spin is prepared into the pure state $\left\vert
0\right\rangle $ by optical pumping.\cite{Harrison:2004uq} A subsequent
$\pi/2$ rotation $\hat{R}_{y}^{e}(\pi/2)$ initializes the electron spin into the
superposition $\left\vert +\right\rangle =(\left\vert 0\right\rangle
+\left\vert 1\right\rangle )/\sqrt{2}$. The nuclear spin is a
partially polarized state $\hat{\rho}_{n}=\hat{I}/2+q_{z}\hat{\sigma}_{z}/2$.
This initial density matrix $\hat{\rho}_{\mathrm{initial}}=\left\vert +\right\rangle
\left\langle +\right\vert \otimes\hat{\rho}_{n}$ coincides with the initial
density matrix $\hat{\rho}_{\mathrm{DQC1}}=\left\vert +\right\rangle
\left\langle +\right\vert \otimes\hat{\rho}_{\mathrm{tar}}$ of the DQC1 model,
where the target qubit state $\hat{\rho}_{\mathrm{tar}}$ also has an arbitrary
polarization, as discussed at the end of Sec. \ref{SEC_DQC1_1}.

\item The two qubits experience a composite evolution (within the
dashed box in Fig.~\ref{G_PROTOCOL}), which consists of a free evolution
$e^{-i\hat{H}\tau}$, a controlled nuclear spin $\pi$ rotation $\tilde{R}%
_{y}^{n}(\pi)=\left\vert 1\right\rangle \left\langle 1\right\vert
\otimes(-i\hat{\sigma}_{y})+\left\vert 0\right\rangle \left\langle
0\right\vert $, an electron spin rotation $\hat{R}_{y}^{e}(\pi)$, another controlled
nuclear spin $\pi$ rotation $\tilde{R}_{y}^{n}(\pi)$, and another free
evolution $e^{-i\hat{H}\tau}$. The equality $\tilde{R}_{y}^{n}(\pi)\hat{R}_{y}%
^{e}(\pi)\tilde{R}_{y}^{n}(\pi)=\hat{R}_{y}^{n}(\pi)\hat{R}_{y}^{e}(\pi)$ shows that this
composite evolution coincides with $\hat{U}_{\mathrm{com}}$ in
Sec.~\ref{SEC_SOLUTION}.

\item A $\pi/2$ rotation $\hat{R}_{y}^{e}(\pi/2)$ is applied to the electron spin,
followed by a measurement of $\hat{Z}$ through optical
methods.\cite{Childress:2006uq, Buckley:2010fk} This measurement estimates%
\begin{align*}
\langle\hat{Z}\rangle &  =\operatorname*{Tr}\hat{Z}\hat{R}_{y}^{e}(\pi/2)\hat
{U}_{\mathrm{com}}\hat{\rho}_{\mathrm{initial}}\hat{U}_{\mathrm{com}}%
^{\dagger}[\hat{R}_{y}^{e}(\pi/2)]^{\dagger}\\
&  =\operatorname*{Tr}\hat{X}e^{-i\hat{H}_{\mathrm{echo}}\tau}\hat{\rho
}_{\mathrm{initial}}e^{i\hat{H}_{\mathrm{echo}}\tau}+O(\eta^{2}).
\end{align*}
Since the evolution $e^{-i\hat{H}_{\mathrm{echo}}\tau}=e^{-i\hat
{H}_{\mathrm{DQC1}}\tau}|_{\theta\rightarrow-A}$ has the same form as the DQC1
model, the average value is
\begin{equation}
\langle\hat{Z}\rangle=\cos(A\tau)+O(\eta^{2}). \label{Zop}%
\end{equation}

\end{enumerate}

The electron spin rotation $\hat{R}_{y}^{e}(\pi/2)$ [$\hat{R}_{y}^{e}(\pi)$] in the
circuit is achieved by a $\pi/2$ pulse ($\pi$ pulse) with the central
frequency $\left\vert D^{\prime}\right\vert $ and the bandwidth $\gg A/2$, so
that both transitions $\left\vert 0,\uparrow\right\rangle \leftrightarrow
\left\vert 1,\uparrow\right\rangle $ and $\left\vert 0,\downarrow\right\rangle
\leftrightarrow\left\vert 1,\downarrow\right\rangle $ are equally excited. The
controlled nuclear spin rotation $\tilde{R}_{y}^{n}(\pi)$ is achieved by a
$\pi$ pulse centered at the resonant frequency $A-g_{N}\mu_{N}B-\delta$ of the
transition $\left\vert 1,\uparrow\right\rangle \rightarrow\left\vert
1,\downarrow\right\rangle $. The duration $\tau$ of the free evolution can be
chosen in the experiment as $\tau>1/A\sim0.1$ $\mathrm{\mu s}$. The electron
spin rotation occurs within a few nanoseconds and hence can be regarded as
instantaneous.\cite{Fuchs:2009fk,Fuchs;NP;2010} However, the controlled
nuclear spin $\pi$ rotation takes $\tau_{n}\sim$ a few microseconds,
comparable to the free evolution time $\tau$. Detailed analysis in appendix A
shows that incorporation of $\tau_{n}$ amounts to replacing the free evolution
time $\tau$ in Eq.~(\ref{Zop}) by the sum $(\tau+\tau_{n})$. For brevity, we
use $\tau$ to denote $(\tau+\tau_{n})$ from now on.

In arriving at Eq.~(\ref{Zop}), we have assumed that all the gate operations in the circuit and
the measurements of $\hat{Z}$ are free of errors. In a realistic experiment,
the most basic errors include the deviation of the nuclear spin rotation
angle from $\pi$ in the controlled $\pi$ rotation $\tilde{R}_{y}^{n}(\pi)$ and
the finite electron spin coherence time $T_{2}^{e}$:

\begin{itemize}
\item Nuclear spin rotation error. The two controlled nuclear spin $\pi$
rotations $\tilde{R}_{y}^{n}(\pi)$ in the quantum estimation circuit
(Fig.~\ref{G_PROTOCOL}) are subjected to random errors, which may come from our limited prior knowledge (which becomes more and more precise after each successive estimation step) about the interaction strength A or other experimental sources. For the actual
rotation angle $(\pi+2\epsilon)$ differing from $\pi$ by an error $2\epsilon$,
the actual controlled rotation $\tilde{R}_{y}^{n}(\pi,\epsilon)=\tilde{R}%
_{y}^{n}(\pi)+\tilde{\delta}_{y}^{n}(\pi)$ differs from the ideal one
$\tilde{R}_{y}^{n}(\pi)$ by%
\begin{equation}
\tilde{\delta}_{y}^{n}(\pi)=\left\vert 1\right\rangle \left\langle
1\right\vert \otimes(-\epsilon+i\frac{\epsilon^{2}}{2}\hat{\sigma}%
_{y})+O(\epsilon^{3}).\nonumber
\end{equation}
For the first controlled rotation being $\tilde{R}_{y}^{n}(\pi,\epsilon_{a})$
and the second controlled rotation being $\tilde{R}_{y}^{n}(\pi,\epsilon_{b})$, the actual quantity estimated by the quantum circuit $M(\tau)$ is
\begin{align}
\langle \hat{Z}_{\epsilon} \rangle =& \left( 1 - \frac{ \langle \epsilon^{2}_{a} \rangle + \langle \epsilon^{2}_{b} \rangle }{2} \right) \cos(A\tau) \nonumber \\
&+ \langle \epsilon_{a}\epsilon_{b} \rangle + \langle \epsilon_{a} \rangle O(\eta) + \langle \epsilon_{b} \rangle O(\eta) + O(\eta^{2}), \nonumber
\end{align}
The first source of error is our ignorance about $A$. In the $k$-th estimation
step, our limited prior knowledge about $A$ (as quantified by the standard
deviation $\Delta_{k-1}$, see Sec.~\ref{SEC_STEPK})
and hence the resonant frequency $A-g_{N}\mu_{N}B-\delta$ of the transition
$\left\vert 1,\uparrow\right\rangle\rightarrow\left\vert 1,\downarrow\right\rangle $
makes it impossible to
construct an exact $\pi$ pulse for this transition. The typical detuning for
this transition is $\Delta_{k-1}$. The typical rotation angle deviates from
the ideal value $\pi$ by an amount $\pi\Delta_{k-1}^{2}/(2\Omega^{2})\sim10^{-3}$, the same order of magnitude as $O(\eta)$, for the Rabi frequency $\Omega=500$~kHz used in our
estimation. Thus every term in the second line of the above equation has the same order of $\sim10^{-6}$, which allows us to replace the second line by
$O(\eta^{2})$. For $\epsilon_{a}$ and $\epsilon_{b}$ being independent, we obtain
\begin{equation}
\langle\hat{Z}_{\epsilon}\rangle=(1-\varepsilon^{2})\cos(A\tau
)+O(\eta^{2}),\nonumber
\end{equation}
where $\varepsilon^{2}=\langle\epsilon_{a}^{2}\rangle=\langle
\epsilon_{b}^{2}\rangle$. For other experimental sources, the errors are
typically random with $\langle\epsilon_{a}\rangle=\langle\epsilon_{b} \rangle=\langle\epsilon_{a}\epsilon_{b}\rangle=0$, so that the above equation
still holds.

\item Electron spin decoherence. The electron spin in the NV center is
subjected to decoherence by the surrounding $^{13}\text{C}$ nuclear spin bath.
The coherence time of the electron spin in the ground state is $T_{2}^{e}%
\sim350$~$\mathrm{\mu s}$ under the natural abundance of the $^{12}\text{C}$
isotope ($98.8\%$), and it is extended to $1.8\text{ ms}$ under the ultrapure
$^{12}\text{C}$ abundance ($99.7\%$) at room temperature.\cite{Gaebel:2006fk,
Balasubramanian:2009uq} By incorporating the electron spin relaxation (with the relaxation time\cite{Boris;PRB;2011} $T_{1}^{e}=5.9$ ms) and decoherence in the
Lindblad form, it is straightforward to show that the quantity estimated by
the quantum circuit is no longer Eq. (\ref{Zop}) but instead
\[
\langle\hat{Z}_{d}\rangle=e^{-2\tau/T_{2}^{e}}\cos\left(  A\tau\right)
+O(\eta^{2}).
\]
\end{itemize}

In summary, in the presence of errors, the quantity estimated by the quantum
circuit in Fig.~\ref{G_PROTOCOL} is given by
\begin{equation}
\langle\hat{Z}\rangle=Q(\tau)\cos(A\tau)+O(\eta^{2}), \label{Zop_REALISTIC}%
\end{equation}
where $Q(\tau)=1-\varepsilon^{2}$ for the nuclear spin rotation error of
magnitude $\varepsilon$ and $Q(\tau)=e^{-2\tau/T_{2}^{e}}$ for a finite
electron spin coherence time $T_{2}^{e}$. In our estimation, we use $B=0.2$~T
so that the correction for the hyperfine interaction $O(\eta^{2})\sim10^{-6}$.

\subsection{Estimation procedure}
\label{SEC_STEPK}

We use $M(\tau)$ to denote the quantum estimation circuit in
Fig.~\ref{G_PROTOCOL}, whose total duration is $2\tau$. A single run of the
circuit $M(\tau)$ returns two outcomes: $+1$ for the electron spin in
the state $\left\vert 0\right\rangle $ or $-1$ for the electron spin in the
state $\left\vert 1\right\rangle $, with corresponding probabilities $p_{\pm
1}=[1\pm\langle\hat{Z}\rangle]/2$. An estimator of the average value
$\langle\hat{Z}\rangle$ [Eq.~(\ref{Zop_REALISTIC})] is obtained by averaging
over the outcomes of repeated running of the circuit. For example, averaging
over $N$ measurements produces $Z$, a single estimator of $\langle\hat{Z}\rangle$.
By the central limit theorem, for relatively large $N$ (e.g.,
$N\gtrsim100$), \textit{this estimator obeys the Gaussian distribution}
$\mathcal{N}(\langle\hat{Z}\rangle,\zeta)$ centered at $\langle\hat
{Z}\rangle$ with a standard deviation $\zeta=1/\sqrt{N}$. Alternatively,
we can also say that \textit{the average value }$\langle\hat{Z}\rangle
$\textit{ obeys the Gaussian distribution} $\mathcal{N}(Z,\zeta)$,
which actually means that the difference $\langle\hat{Z}\rangle-Z$ obeys
the Gaussian distribution $\mathcal{N}(0,\zeta)$.

The estimation begins with a prior knowledge of the hyperfine interaction
strength $A$. It is quantified by a Gaussian distribution $\mathcal{N}%
(A_{0},\Delta_{0})$ centered at $A_{0}$ with a relatively large standard
deviation $\Delta_{0}$, which quantifies our ignorance about $A$. This prior
knowledge tells us, with a 95\% confidence, that $A$ lies within the interval
$[A_{0}-1.96\Delta_{0},A_{0}+1.96\Delta_{0}]$. From the prior knowledge
$\mathcal{N}(A_{0},\Delta_{0})$, we construct the quantum circuit $M(\tau
_{1})$ for the first estimation, which provides a new knowledge about $A$, as
quantified by a Gaussian distribution $\mathcal{N}(\bar{A}_{1},\bar{\Delta
}_{1})$. Through the Bayesian inference, this new knowledge is combined with
the prior knowledge to produce an updated knowledge about $A$, quantified by a
Gaussian distribution $\mathcal{N}(A_{1},\Delta_{1})$ with a smaller standard
deviation $\Delta_{1}<\Delta_{0}$. Therefore, the first estimation step
refines our knowledge about $A$ from $\mathcal{N}(A_{0},\Delta_{0})$ to
$\mathcal{N}(A_{1},\Delta_{1})$ (with $\Delta_{1}<\Delta_{0}$), which in turn
serves as the prior knowledge of the next estimation step. By iterating this
procedure, the standard deviation of the Gaussian distribution quantifying our
ignorance about $A$ would decrease successively as $\Delta_{0}>\Delta
_{1}>\Delta_{2}>\cdots$. The iteration is stopped at the $K$-th step when the
desired standard deviation $\Delta_{\mathrm{desire}}$ is achieved: $\Delta
_{K}\leq\Delta_{\mathrm{desire}}$. Below, we describe the above estimation
procedures in more detail.

\subsubsection{Gaining knowledge about A from measurements}

In the $k$-th estimation step $(k=1,2,\cdots)$, the prior knowledge about the hyperfine interaction strength $A$ is quantified by the Gaussian distribution $N(A_{k-1},\Delta_{k-1})$. Suppose that
$\tau_{k}$ has been properly chosen (to be discussed shortly). By running the
circuit $M(\tau_{k})$ for a relatively large number $N_{k}$ ($\gtrsim100$) of
times, we obtain an estimator $Z_{k}$ of $\langle\hat{Z}\rangle_{k}\equiv
Q(\tau_{k})\cos(A\tau_{k})+O(\eta^{2})$ with a standard deviation
$\zeta_{k}=1/\sqrt{N_{k}}$. This knowledge tells us that $\langle\hat
{Z}\rangle_{k}$ obeys the Gaussian distribution $\mathcal{N}(Z_{k}%
,\zeta_{k})$. We need to convert this distribution of $\langle\hat
{Z}\rangle_{k}$ to a distribution of $A$. For a general $\tau_{k}$, the
relation between $\langle\hat{Z}\rangle_{k}$ and $A$ is nonlinear and the
conversion from $\langle\hat{Z}\rangle_{k}$ to $A$ results in a non-Gaussian
distribution of $A$, with a characteristic width%
\[
\frac{\zeta_{k}}{|\partial\langle\hat{Z}\rangle_{k}/\partial A|}%
=\frac{\zeta_{k}}{Q(\tau_{k})\tau_{k}|\sin(A\tau_{k})|}.
\]
Now we determine $\tau_{k}$ according to two requirements:

\begin{enumerate}
\item The distribution of $A$ should be Gaussian (i.e., the relation between
$\langle\hat{Z}\rangle_{k}$ and $A$ should be linear), so that analytical
results can be obtained. Based on our prior knowledge $\mathcal{N}%
(A_{k-1},\Delta_{k-1})$ about $A$, the conditions $A_{k-1}\tau_{k}=\pi/2+2\pi
\times\operatorname{integer}$ and $\Delta_{k-1}\tau_{k}\ll1$ enable the Taylor
expansion $\langle\hat{Z}\rangle_{k}=(A_{k-1}-A)Q(\tau_{k})\tau_{k}+\delta
_{k}+O(\eta^{2})$ with $\delta_{k} \approx Q(\tau_{k})(\Delta_{k-1}\tau_{k}%
)^{3}/6$. For $\delta_{k},O(\eta^{2})\ll\zeta_{k},|\langle\hat{Z}%
\rangle_{k}|$, the correction terms $\delta_{k}+O(\eta^{2})$ can be safely
dropped, so that the relation between $\langle\hat{Z}\rangle_{k}$ and $A$
becomes linear and the distribution of $A$ becomes Gaussian $\mathcal{N}%
(\bar{A}_{k},\bar{\Delta}_{k})$ with
\begin{subequations}
\label{A1BAR_TAOkBAR}%
\begin{align}
\bar{A}_{k}  &  =A_{k-1}-\frac{Z_{k}}{Q(\tau_{k})\tau_{k}},\label{A1BAR}\\
\bar{\Delta}_{k}  &  =\frac{\zeta_{k}}{Q(\tau_{k})\tau_{k}}=\frac
{1}{Q(\tau_{k})\tau_{k}\sqrt{N_{k}}}. \label{DELTAkBAR}%
\end{align}
The distribution $\mathcal{N}(\bar{A}_{k},\bar{\Delta}_{k})$ of $A$ tells us,
with a 95\% confidence, that $A$ lies in the interval $[\bar{A}_{k}%
-1.96\bar{\Delta}_{k},\bar{A}_{k}+1.96\bar{\Delta}_{k}]$.

\item For maximal precision of the estimation, the standard deviation
$\bar{\Delta}_{k}$ should be minimized, i.e., $Q(\tau_{k})\tau_{k}$ should be maximized.
\end{subequations}
\end{enumerate}

Eq.~(\ref{DELTAkBAR}) shows that the standard deviation $\bar{\Delta}%
_{k}$ of the measurement of $A$ is equal to the standard deviation
$\zeta_{k}=1/\sqrt{N_{k}}$ of the measurement of $\langle\hat{Z}%
\rangle_{k}$ divided by $Q(\tau_{k})\tau_{k}$:

\begin{itemize}
\item For $Q(\tau_{k})=1$ (i.e., no errors), the standard deviation
$\bar{\Delta}_{k}$ is reduced upon the increase of $\tau_{k}$, which can be
interpreted as a repetition of the circuit operations (as enclosed in the
dashed box in Fig.~\ref{G_PROTOCOL}) before the measurement is made. This is
equivalent to a multiround protocol suggested by Giovannetti \textit{et
al}.\cite{Giovannetti;PRL;2006}. Therefore, the dependence $\bar{\Delta}%
_{k}\propto1/\tau_{k}$ implies the QML.

\item The standard deviation $\bar{\Delta}_{k}$ is reduced upon the increase
of $N_{k}$. The dependence $\bar{\Delta}_{k}\propto1/\sqrt{N_{k}}$ implies the SQL.
\end{itemize}

In summary, for optimal performance, we should first choose $\zeta_{k}$
(or equivalently $N_{k}$) subjected to the constraint
\begin{equation}
O(\eta^{2})\ll\zeta_{k}\ll1 \label{CONSTRAINT_NK}%
\end{equation}
and then choose $\tau_{k}$ to maximize $Q(\tau_{k})\tau_{k}$, subjected to the
constraints
\begin{subequations}
\label{CONSTRAINT_TAOK}%
\begin{align}
& A_{k-1}\tau_{k}  =\frac{\pi}{2}+2m_{k}\pi,\\
& \frac{(\Delta_{k-1}\tau_{k})^{3}}{6} \ll \zeta_{k}, \label{CONSTRAINT_B} \\
& Q(\tau_{k})\Delta_{k-1}\tau_{k} \gg O(\eta^{2}), \label{CONSTRAINT_C}
\end{align}
where $m_{k}\in\mathbb{Z}$ and $O(\eta^{2})\sim10^{-6}$ for $B=0.2$ T. The constraint $\zeta_{k}%
\ll1$ ensures the validity of our Gaussian distribution assumption for
$\langle\hat{Z}\rangle_{k}$, while other constraints ensure the validity of
the formula $\langle\hat{Z}\rangle_{k}\approx(A_{k-1}-A)Q(\tau_{k}%
)\tau_{k}$. The error of the linear expansion can be dropped if $\delta_{k} \ll \zeta_{k}$, which gives Eq.~(\ref{CONSTRAINT_B}) with $Q(\tau_{k}) \leq 1$. Eq.~(\ref{CONSTRAINT_C}) denotes the condition to drop $O(\eta^{2})$ in $\langle \hat{Z} \rangle_{k}$. Note that the constraints [Eqs.~(\ref{CONSTRAINT_TAOK})] on
$\tau_{k}$ have no solution under certain conditions, e.g., when $Q(\tau
_{k})\lesssim O(\eta^{2})/(\zeta_{k})^{1/3}$. Therefore, for more flexible
choice of $\tau_{k}$, the standard deviation $\zeta_{k}$ of the
measurement of $\langle\hat{Z}\rangle_{k}$ should not be too small.

\subsubsection{Combining new knowledge with prior knowledge}

In the previous subsection, we have spent $N_{k}$ runs of the circuit
$M(\tau_{k})$ to obtain the new knowledge $\mathcal{N}(\bar{A}_{k},\bar
{\Delta}_{k})$ about $A$. To make use of the resources spent in obtaining the
prior knowledge $\mathcal{N}(A_{k-1},\Delta_{k-1})$, we use the Bayesian
inference, which combines our new knowledge $\mathcal{N}(\bar{A}_{k}%
,\bar{\Delta}_{k})$ with the prior knowledge $\mathcal{N}(A_{k-1},\Delta_{k-1})$.
It gives an updated Gaussian distribution $\mathcal{N}(A_{k},\Delta_{k})$
centered at%
\end{subequations}
\begin{subequations}
\begin{equation}
A_{k}=\frac{A_{k-1}/\Delta_{k-1}^{2}+\bar{A}_{k}/\bar{\Delta}_{k}^{2}}%
{1/\Delta_{k-1}^{2}+1/\bar{\Delta}_{k}^{2}} \label{AK}%
\end{equation}
(which is a weighted average of $A_{k-1}$ with weight $1/\Delta_{k-1}^{2}$ and
$\bar{A}_{k}$ with weight $1/\bar{\Delta}_{k}^{2}$) with a standard deviation
$\Delta_{k}$ determined by%
\begin{equation}
\frac{1}{\Delta_{k}^{2}}=\frac{1}{\Delta_{k-1}^{2}}+\frac{1}{\bar{\Delta}%
_{k}^{2}}. \label{DELTAK}%
\end{equation}
This updated knowledge $\mathcal{N}(A_{k},\Delta_{k})$ tells us, with a 95\%
confidence, that $A$ lies in the refined interval $[A_{k}-1.96\Delta_{k}%
,A_{k}+1.96\Delta_{k}]$. The inequalities $\Delta_{k}<\Delta_{k-1}$ and
$\Delta_{k}<\bar{\Delta}_{k}$ reveal that the combination of $\mathcal{N}%
(A_{k-1},\Delta_{k-1})$ and $\mathcal{N}(\bar{A}_{k},\bar{\Delta}_{k})$ gives us a
more precise knowledge about $A$.

For very accurate measurement compared with the prior knowledge, i.e.,
$\bar{\Delta}_{k}\ll\Delta_{k-1}$, Eqs.~(\ref{AK}) and (\ref{DELTAK}) reduce to
$A_{k}\approx\bar{A}_{k}$ and $\Delta_{k}\approx\bar{\Delta}_{k}$, suggesting
that the updated knowledge is dominated by the measurement. By contrast, for
inaccurate measurement $\bar{\Delta}_{k}\gg\Delta_{k-1}$, the updated knowledge
$A_{k}\approx A_{k-1}$ and $\Delta_{k}\approx\Delta_{k-1}$ is dominated by the
prior knowledge.

\subsection{Ideal case: approaching quantum metrology limit}
\label{SEC_IDEAL}

In this subsection, we demonstrates the QML scaling of our estimation protocol
in the ideal case, i.e., in the absence of any errors (e.g., operation errors,
relaxation, and decoherence). For simplicity, we assume that in each
estimation step, we run the quantum circuit for the same number of times
$N_{1}=N_{2}=\cdots\equiv N$, corresponding to $\zeta_{1}=\zeta_{2}%
=\cdots\equiv\zeta\equiv1/\sqrt{N}$.

Up to the $K$-th estimation step, the total duration of the our estimation
process (identified as the total amount of resources spent) is
\end{subequations}
\[
R_{K}=N\sum_{k=1}^{K}2\tau_{k}\equiv N\tau_{K}^{\text{tot}}.
\]
To see the scaling of the precision $1/\Delta_{K}^{2}$ with respect to $R_{K}%
$, we take the first estimation step as a reference. Further, we take
$\Delta_{0}=\infty$ to exclude the contribution from the prior knowledge
$\mathcal{N}(A_{0},\Delta_{0})$, so that all our knowledge about $A$ comes
from the resources $R_{K}$ spent in our protocol. Then, the QML limit
$\Delta_{K,\mathrm{QML}}$ is defined by $\Delta_{K,\mathrm{QML}}/\Delta
_{1}\equiv1/(R_{K}/R_{1})$, while the SQL limit $\Delta_{K,\mathrm{SQL}}$ is
defined by $\Delta_{K,\mathrm{SQL}}/\Delta_{1}\equiv1/\sqrt{R_{K}/R_{1}}$.
Using $R_{1}=2N\tau_{1}$ and $\Delta_{1}=1/(\tau_{1}\sqrt{N})$, we
obtain%
\begin{align}
\frac{1}{\Delta_{K,\mathrm{QML}}^{2}}  &  =N\left(  \sum_{k=1}^{K}\tau
_{k}\right)  ^{2},\label{DELTAK_QML}\\
\frac{1}{\Delta_{K,\mathrm{SQL}}^{2}}  &  =N\tau_{1}\sum_{k=1}^{K}\tau
_{k},\label{DELTAK_SQL} \\
\frac{1}{\Delta_{K}^{2}}  &  =N\sum_{k=1}^{K}\tau_{k}^{2}.\nonumber
\end{align}

First, we compare $\Delta_{K}$ with the QML limit $\Delta_{K,\mathrm{QML}}$
and the SQL limit $\Delta_{K,\mathrm{SQL}}$ and discuss the condition for
approaching the QML:

\begin{enumerate}
\item The inequality $\Delta_{K}>\Delta_{K,\mathrm{QML}}$ can be readily
verified. This manifests the QML precision $1/\Delta_{K,\mathrm{QML}}^{2}$ as
the upper precision bound. To achieve the QML, $\{\tau_{k}\}$ should satisfy
$\tau_{K}\gg\tau_{K-1}\gg\cdots\gg\tau_{1}$, so that the total amount of
resources is dominated by the final estimation step and hence $\Delta
_{K}\approx\Delta_{K,\mathrm{QML}}\approx1/(\tau_{K}\sqrt{N})$. This condition
is equivalent to a dramatic reduction of the standard deviation of the
measurement for each successive estimation step: $\bar{\Delta}_{K}\ll
\bar{\Delta}_{K-1}\ll\dots\ll\bar{\Delta}_{1}$. This ensures that in each
estimation step (say, the $k$-th step), the standard deviation of the
estimation, $\Delta_{k}\approx\bar{\Delta}_{k}\approx1/(\tau_{k}\sqrt{N})$, is
dominated by the standard deviation $\bar{\Delta}_{k}$ of the measurement
instead of the standard deviation $\Delta_{k-1}\approx\bar{\Delta}_{k-1}$ of
the prior knowledge [cf. Eq.~(\ref{DELTAK})]. The condition $\tau_{K}\gg
\tau_{K-1}\gg\cdots\gg\tau_{1}$ is also equivalent to
\begin{equation}
\Delta_{k-1}\tau_{k}\gg\zeta, \label{QML_CONDITION}%
\end{equation}
since $\Delta_{k-1}\tau_{k}\approx\bar{\Delta}_{k-1}\tau_{k}=(\tau_{k}%
/\tau_{k-1})\zeta$.

\item For $\tau_{1}=\tau_{2}=\cdots=\tau_{K}$, the precision $1/\Delta_{K}%
^{2}=NK\tau_{1}^{2}$ coincides with the SQL precision $1/\Delta
_{K,\mathrm{SQL}}^{2}$ since in this case our protocol reduces to simple
repetition of the same quantum circuit $M(\tau_{1})$.
\end{enumerate}

\begin{figure}[ptb]
\includegraphics[width=\columnwidth,clip]{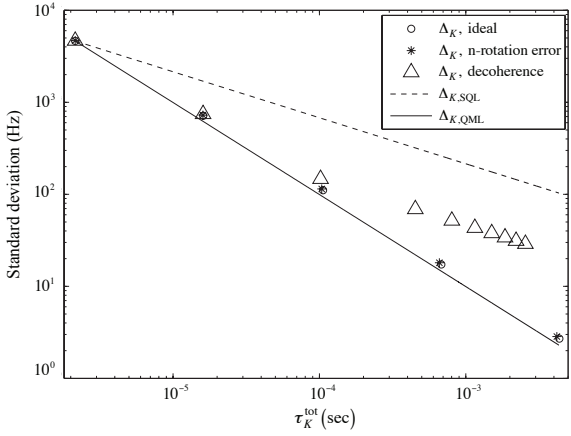}\caption{
Comparision of the standard deviation $\Delta_{K}$ of our protocol with the
QML limit $\Delta_{K,\mathrm{QML}}$ (solid line) and the SQL limit
$\Delta_{K,\mathrm{SQL}}$ (dashed line). How to choose $\tau_{k}$ is explained in the main text.}%
\label{G_PRECISION}%
\end{figure}

Then we give the best choice $\{\tau_{k}^{\mathrm{ideal}}\}$ satisfying the QML condition Eq.~(\ref{QML_CONDITION}) for the ideal case according to the description in Sec.~\ref{SEC_STEPK}. We choose $\{\tau_{k}^{\mathrm{ideal}}\}$ by taking the largest $m_{k}$ such that $\Delta_{k-1}\tau_{k} \approx c$ at every step, where $c$ is a constant satisfying $c \gg \zeta$ and $c^{3} \ll 6\zeta$. Then $\{\tau_{k}^{\mathrm{ideal}}\}$ automatically satisfies the QML condition Eq.~(\ref{QML_CONDITION}) and the linear expansion condition Eq.~(\ref{CONSTRAINT_B}).
From $\Delta_{k-1}\tau_{k} \approx c$, we have $\Delta_{k}\approx\bar{\Delta}%
_{k}\approx(\zeta/c)^{k}\Delta_{0}$, i.e., the standard deviation
$\Delta_{k}\approx\bar{\Delta}_{k}$ is dramatically reduced by each successive
estimation step. We also have $\tau_{k}^{\mathrm{ideal}} \approx (c/\zeta)^{k}
\tau_{0}$ (with $\tau_{0}$ defined through $\Delta_{0}\equiv\zeta/\tau_{0}$),
i.e., an exponential increase of $\tau_{k}^{\mathrm{ideal}}$ with $k$. Note that,
for $B=0.2$ T, we have $O(\eta^{2})\sim10^{-6}$. Therefore $\zeta$ can be as
small as $\sim10^{-5}$.

Finally we provide a numerical simulation for the estimation process. The
parameters for the simulation are $A=3.06$~MHz, $B=0.2$~T, $N=1000$,
corresponding to $\zeta\approx0.03$. We take
$c=0.2$, which satisfies $c \gg \zeta$ and $c^{3} \ll 6\zeta$. The prior knowledge is $A_{0}=3.03$~MHz
with a standard deviation $\Delta_{0}=0.03$~MHz, which has been reported by a
previous experiment.\cite{Felton;PRB;2009} Each controlled nuclear spin $\pi$
rotation uses a $1$-$\mathrm{\mu}$\textrm{s }square pulse with the Rabi frequency
$\Omega=500$~kHz. The electron spin rotations are regarded as instantaneous,
as mentioned at the end of Sec. \ref{SEC_CIRCUIT}. In Fig. \ref{G_PRECISION},
the proximity of $\Delta_{K}$ (circles) to $\Delta_{K,\mathrm{QML}}$ (solid
line) confirms the QML scaling of the estimation.

\subsection{Realistic case: surpassing standard quantum limit}

In this subsection, we take into account the nuclear spin rotation error and
electron spin decoherence and discuss the optimal choice
of $\{\tau_{k}\}$ and the resulting precision%
\[
\frac{1}{\Delta_{K}^{2}}=N\sum_{k=1}^{K}[Q(\tau_{k})\tau_{k}]^{2}%
\]
of the estimation, derived from Eq.~(\ref{DELTAkBAR}) and (\ref{DELTAK}):

\begin{itemize}
\item Nuclear spin rotation error $Q(\tau)=1-\varepsilon^{2}\equiv Q$.
This error is equivalent to an increase of $\zeta$ to $\tilde{\zeta}%
\equiv\zeta/Q$. Then QML condition Eq.~(\ref{QML_CONDITION}) becomes $\Delta_{k-1}\tau_{k} \gg \tilde{\zeta}$.
For a general $Q$ that is not too small (i.e., $1 \geq Q \gg \zeta$),
the conclusion in the ideal case remains valid with $\zeta\rightarrow\tilde{\zeta}$,
i.e., $\{\tau_{k}\}$ is chosen as $\tau_{k}\approx(c/\tilde{\zeta})^{k}(\tau_{0}/Q)$, where $c$ is a constant subjected to $c \gg \tilde{\Delta}_{Z}$ and $c^{3} \ll 6\tilde{\Delta}_{Z}$. In the simulation, we consider a typical error $\varepsilon=0.1$ (corresponding to $\sim3\%$ error in the rotation angle).
Then we have $Q\approx1$, and this allows us to set $c=0.2$, the same value with the ideal case. As a result, we can choose $\tau_{k} \approx \tau_{k}^{\mathrm{ideal}}$ and $\Delta_{K}$
is nearly the same as the ideal case. Therefore, the QML scaling is preserved for
the realistic nuclear spin rotation error, as confirmed by the nearly complete
coincidence between $\Delta_{K}$ (stars) and $\Delta_{K,\mathrm{QML}}$ (solid
line) in Fig.~\ref{G_PRECISION}.

\item Electron spin decohence $Q(\tau)=e^{-2\tau/T_{2}^{e}}$. According to
Sec.~\ref{SEC_STEPK}, we should choose $\tau_{k}$ to maximize $Q(\tau_{k}%
)\tau_{k}$, subjected to the constraints in Eqs.~(\ref{CONSTRAINT_TAOK}).
We use $\Delta_{k-1}\tau_{k} \approx c=0.2$ in the simulation. In the presence of the electron spin decoherence, $Q(\tau)$ decreases as $\tau$ increases. Thus the QML condition $c \gg \zeta/Q(\tau_{k})$ is no longer valid at some point. This is why $\Delta_{k}$ starts to deviate from the QML line at $k=3$ in Fig.~\ref{G_PRECISION}. Note that the estimation of $k=3$ still surpasses the SQL. The maximum of $Q(\tau)\tau$ occurs at $\tau=T^{e}_{2}/2$, meaning that the standard deviation $\bar{\Delta}_{k}$ of the quantum circuit $M(\tau_{k})$ is the smallest when $\tau_{k} \approx T^{e}_{2}/2$. Further increase of $\tau_{k}$ makes the precision of $M(\tau_{k})$ worse. Once $\tau_{k}$ reaches $\tau_{k} \approx T^{e}_{2}/2$ at $k=k_{c}$, the estimation for $k>k_{c}$ is performed with $\tau_{k} = \tau_{k_c}$. Therefore, for $K=k_{c}+\tilde{K}$, further
estimation steps beyond $k_{c}$ (i.e., $k=k_{c}+1,\cdots,k_{c}+\tilde{K}$) increases the precision $1/\Delta_{K}^{2}$ by the SQL trend:
\[
\frac{1}{\Delta_{k_{c}+\tilde{K}}^{2}}-\frac{1}{\Delta_{k_{c}}^{2}}\approx
N\tilde{K}(T_{2}^{e}/2)^{2}.
\]
For $T_{2}^{e}=350$ $\mathrm{\mu s}$, we have $k_{c}=4$. Fig.~\ref{G_PRECISION} shows that $\Delta_{K}$ surpasses the SQL for $K<4$, while
it decreases parallel to the SQL for $K\geq4$.
\end{itemize}

\section{Conclusions}

\label{SEC_CONCLUSION}

We have proposed an efficient quantum measurement protocol to estimate the
hyperfine interaction between the electron spin and the $^{15}$N nuclear spin
in the NV center. The essential idea of our protocol is the combination of the
DQC1 parameter estimation\cite{Boixo;PRA;2008} with the spin-echo technique.
The spin echo eliminates the independent dynamics of the electron spin and the
nuclear spin in the DQC1 model, but keeps the dynamics due to their
interactions, whose strength is to be estimated. This protocol does not require
the preparation of the nuclear spin state. We quantify the resources $R$ as
the total duration $\sum\tau$ of the estimation process. In the absence of any
errors, the precision $1/\Delta^{2}$ (with $\Delta$ being the standard
deviation) of the estimation approaches the quantum metrology limit (QML)
$1/\Delta_{\mathrm{QML}}^{2}=O(R^{2})$. This QML scaling is robust against the
typical nuclear spin rotation error in realistic experimental conditions. In the presence of electron spin decoherence, the precision $1/\Delta^{2}$ keeps its QML scaling when $\tau \ll T^{e}_{2}/2$.
Once $\tau$ becomes close to $T^{e}_{2}$ further estimation steps
increase the precision $1/\Delta^{2}$ according to the scaling
$1/\Delta_{\mathrm{SQL}}^{2}=O(R)$ of the standard quantum limit (SQL).
Due to the QML scaling in the initial stage, the overall precision still surpasses the
SQL. We expect that this method can be applied to other
solid state systems such as quantum dots or cold atoms to measure the
interaction between two spins.

\begin{acknowledgments}
This research was supported by the U. S. Army Research Office under contract
number ARO-MURI W911NF-08-2-0032. The authors are grateful to David~M.~Toyli  and Jiangfeng Du for helpful discussions.
\end{acknowledgments}

\appendix

\section{Accounting for finite duration of controlled nuclear spin rotation}

In this section, we assume that each of the two controlled nuclear spin $\pi$
rotation in the quantum protocol (Fig.~\ref{G_PROTOCOL}) is driven by a square
$\pi$ pulse with a duration $\tau_{n}\sim1\ \mathrm{\mu s}$ and prove that
inclusion of this finite duration amounts to a trivial renormalization
$\tau\rightarrow\tau+\tau_{n}$ in Eq.~(\ref{Zop}).

In Fig.~\ref{G_PROTOCOL}, the initial state $\hat{\rho}_{\mathrm{initial}%
}=\left\vert +\right\rangle \left\langle +\right\vert \otimes\hat{\rho}_{n}$
is prepared at $t=-\tau-\tau_{n}$. The first free evolution $e^{-i\hat{H}\tau
}$ occurs during $t\in\lbrack-\tau-\tau_{n},-\tau_{n}]$, followed by a
controlled nuclear spin $\pi$ rotation during $t\in\lbrack-\tau_{n},0]$. A
fast electron spin $\pi$ rotation is applied at $t=0$, another
controlled nuclear spin $\pi$ rotation during $t\in\lbrack0,\tau_{n}]$, and
another free evolution $e^{-i\hat{H}\tau}$ during $t\in\lbrack\tau_{n}%
,\tau+\tau_{n}]$.

First we calculate the evolution operator driven by a square $\pi$ pulse
applied during $t\in\lbrack t_{1},t_{2}]$, with a central frequency
$\omega=A-g_{N}\mu_{N}B-\delta$ (where $+\delta$ is the energy correction to
$\left\vert 1,\uparrow\right\rangle $ by the off-diagonal hyperfine
interaction) resonant with the transition $\left\vert 1,\uparrow\right\rangle
\rightarrow\left\vert 1,\downarrow\right\rangle $. During this pulse, the
Hamiltonian $\hat{H}(t)=\hat{H}+\hat{V}(t)$ of the electron-nuclear spin
qubits acquires an additional term
\begin{equation}
\hat{V}(t)=\frac{i\Omega_{R}}{2}(e^{-i\omega t}\left\vert 1,\downarrow
\right\rangle \left\langle 1,\uparrow\right\vert -e^{i\omega t}\left\vert
1,\uparrow\right\rangle \left\langle 1,\downarrow\right\vert ),\nonumber
\end{equation}
with a constant Rabi frequency $\Omega_{R}=\pi/(t_{2}-t_{1})$. With the aid of
the interaction picture $|\Psi_{\mathrm{I}}(t)\rangle\equiv e^{i\hat{H}t}%
|\Psi(t)\rangle$, the evolution operator $\hat{U}_{V}(t_{2},t_{1})$ during
$t\in\lbrack t_{1},t_{2}]$ can be calculated straightforwardly as $\hat{U}%
_{V}(t_{2},t_{1})=e^{-i\hat{H}t_{2}}e^{-i\hat{H}_{\mathrm{I}}(t_{2}-t_{1}%
)}e^{i\hat{H}t_{1}}$, where $\hat{H}_{\mathrm{I}}(t)\equiv e^{i\hat{H}t}%
\hat{V}(t)e^{-i\hat{H}t}$. Similar to the discussions in Sec.~\ref{SEC_DQC1_2}%
, we have $\hat{H}_{\mathrm{I}}(t)=(\Omega_{R}/2)\left\vert 1\right\rangle
\left\langle 1\right\vert \otimes\hat{\sigma}_{y}+O(\Omega_{R}\eta)$, where
$\eta\sim10^{-3}$ for the external magnetic field $B=0.2$ T used in our
estimation. Therefore, the evolution $e^{-i\hat{H}_{\mathrm{I}}(t_{2}-t_{1}%
)}\approx\tilde{R}_{y}^{n}(\pi)$ coincides with the instantaneous controlled
rotation and hence
\[
\hat{U}_{V}(t_{2},t_{1})=e^{-i\hat{H}t_{2}}\tilde{R}_{y}^{n}(\pi)e^{i\hat
{H}t_{1}}.
\]
With the aid of this result, it can be readily checked that the evolution
operator for the composite evolution (as enclosed by the dashed box) in Fig.
\ref{G_PROTOCOL} is equal to $\hat{U}_{\mathrm{com}}|_{\tau\rightarrow
(\tau+\tau_{n})}$. Therefore, inclusion of the finite duration $\tau_{n}$ of
the controlled nuclear spin rotation amounts to replacing $\tau$ with
$(\tau+\tau_{n})$ in Eq.~(\ref{Zop}). Note that the nuclear spin relaxation
time and decoherence time $\gtrsim1$~ms are much longer than $\tau_{n}\sim1$ $\mathrm{\mu s}$ and hence have negligible influence on this result.\cite{Neumann;Science;2010,
Fuchs:2011fk}

\bibliographystyle{apsrev4-1}

\end{document}